# Sleeping Giants - Activating Dormant Java Deserialization Gadget Chains through Stealthy Code Changes

Bruno Kreyssig
bruno.kreyssig@cs.umu.se
Umeå University
Umeå, Sweden

Sabine Houy
sabine.houy@cs.umu.se
Umeå University
Umeå, Sweden

Timothée Riom
timothee.riom@cs.umu.se
Umeå University
Umeå, Sweden

Alexandre Bartel
alexandre.bartel@cs.umu.se
Umeå University
Umeå, Sweden

## Abstract

Java deserialization gadget chains are a well-researched critical software weakness. The vast majority of known gadget chains rely on gadgets from software dependencies. Furthermore, it has been shown that small code changes in dependencies have enabled these gadget chains. This makes gadget chain detection a purely reactive endeavor. Even if one dependency's deployment pipeline employs gadget chain detection, a gadget chain can still result from gadgets in other dependencies. In this work, we assess how likely small code changes are to enable a gadget chain. These changes could either be accidental or intentional as part of a supply chain attack. Specifically, we show that class serializability is a strongly fluctuating property over a dependency's evolution. Then, we investigate three change patterns by which an attacker could stealthily introduce gadgets into a dependency. We apply these patterns to 533 dependencies and run three state-of-the-art gadget chain detectors both on the original and the modified dependencies. The tools detect that applying the modification patterns can activate/inject gadget chains in *26.08%* of the dependencies we selected. Finally, we verify the newly detected chains. As such, we identify dormant gadget chains in 53 dependencies that could be added through minor code modifications. This both shows that Java deserialization gadget chains are a broad liability to software and proves dormant gadget chains as a lucrative supply chain attack vector.

## 1 Introduction

The native Java serialization API has a long history of being errorprone and vulnerable. By now it has been ten years since *Frohoff* [16] first addressed the inherent security cost induced to Java applications by the *Serializable API*. Specifically, being able to exploit the object-oriented nature of Java's native serialization in code reuse attacks, i.e., gadget chains, has garnered much attention from the research community. Starting with *Serianalyzer* [2] and *GadgetInspector* [21], numerous gadget chain detection tools [5–8, 10, 29, 32, 34, 36, 37, 51, 54, 60, 67] and remediation strategies [11, 12, 52, 70] have been developed.

Yet, despite substantial efforts and initiatives to replace the *Serializable API* with a more secure data-driven mechanism [1, 17–19], it seems to be here to stay. For instance, *Android's* inter-app communication continues to rely on native Java serialization [29] and since 2020, 55 new critical vulnerabilities were published in the *National Vulnerability Database*[1] [40]. All the while, *Ysoserial* [15] – the point of reference for Java deserialization gadget chains – has grown stale (last gadget chain added in February 2021). With continuous software evolution, *Ysoserial* is loosing relevance regarding gadget chains in current software. It has been shown that the occurrence of deserialization gadget chains is related to minor code changes [27, 55], such as adding the *Serializable* interface to a class. Over a Java dependency's evolution, these small changes make a gadget chain appear or disappear. Thus, if a gadget chain detection tool fails to find a gadget chain in the latest version of a dependency, it is no guarantee that gadget chains will not surface in the future. Neither that it may already exist through a combination of dependencies. This makes gadget chain detection in its current state a reactive process. In this work, instead of asking whether a dependency contains gadget chains, we analyze how close a dependency is to containing a gadget chain.

First, we highlight the volatility of Java class serializability over a dependency's evolution by downloading 1 475 widely used dependencies from *Maven*, constituting 111 275 versions. We compare how many classes are serializable in each version and whether changes in serializability occur through direct addition of the serializable interface or transitively through a supertype. In particular, the latter case provides valuable insights into the use of gadget chains in a supply chain attack. While it may be obvious in code reviews that a concrete class containing sensitive method calls should not become serializable, a maintainer could miss that making an interface serializable also impacts subtypes.

This is one of the three modification patterns we investigate, through which either an attacker could stealthily inject or a developer unintentionally add gadgets into a dependency. We apply these changes to the 533 dependencies from the dataset above, which contain at least one serializable class. On both the original and modified dependencies, we run three recent representative gadget chain detection tools: *Tabby* [10], *Crystallizer* [60] and *AndroChain* [29]. This allows us to map gadget chains reported in the modified version, but not the original, as related to the applied modifications. As such, we find that for *26.08%* of the dependencies, the modifications lead to more gadget chain detections. Then, through manual analysis, we confirm dormant gadget chains in 53 of these dependencies. *49.06%* of the true positives required only one of our modification patterns.

---
[1] Using the *Common Weakness* identifier CWE-502, 'Java' as a keyword and a CVSS V3 score of ≥ 9.

Our work highlights that even if Java deserialization gadget chains are not a liability to software dependencies in their current state [29], in many cases, it is feasible to activate these dormant gadget chains either by accident or through a targeted supply chain attack. To make matters worse, maintainers have no control over gadgets from other dependencies, which can be leveraged as a deserialization gadget chain in conjunction with their artifact. It implies the necessity to be aware of partial (dormant) gadget chains so one can remediate these weaknesses before they become a vulnerability. Our main contributions are:

- A study of Java serializability usage over dependency evolution. This also leads us to define four dependency datasets as a basis for future research on Java gadget chains.
- The concept of *dormant gadget chains* as a novel supply chain attack vector (and technical debt) with three attack patterns by which gadgets can be injected into dependencies.
- A feasibility analysis of this attack on 533 dependencies, using three gadget chain detection tools to locate injected chains and manual analysis for verification.
- A ground truth of 53 dependencies with dormant gadget chains, verified through manual analysis.

## 2 Background

### 2.1 Java Deserialization Gadget Chains

A Java deserialization gadget chain describes a sequence of method calls leading from a deserialization entry point to a security-sensitive method. To be clear, the term *entry point* is overloaded. It first represents an insecure *ObjectInputStream* calling readObject() which triggers deserialization of an arbitrary object. For instance, in CVE-2024-45772 [39] (see Listing 1), *Apache Lucene*'s HTTP client[2] implementation would, on receiving a bad response code, attempt to recover the cause from the failed response. One could assume that the object being reconstructed from the stream at line 5 must be of the type *Throwable* to not violate the cast. However, the entire reconstruction, i.e., deserialization, occurs before the cast. This is precisely where a gadget chain payload is triggered.

Research [26, 29, 47] and recent vulnerabilities (e.g., Listing 1) show that such entry points continue to exist in software. Given the existence of deserialization entry points, we shift our attention for the remainder of this work to entry points for deserialization gadget chains. Serializable Java objects may override any of the default deserialization methods `readObject`, `readResolve`, and `readObjectNoData` to implement custom deserialization logic. A

---
[2]https://github.com/apache/lucene/commit/b4b153f64fb0420c6e28ddea4c442e119458756c

```
1  protected void throwKnownError(HttpResponse res, StatusLine s){
2    ObjectInputStream in = null;
3    in = new ObjectInputStream(res.getEntity().getContent());
4    try {
5      Throwable t = (Throwable) in.readObject();
6    } catch (Throwable th) { /*...*/ }
7  }
```

Listing 1: CVE-2024-45772 [39]: insecure deserialization entry point in *Apache Lucene*'s `HttpClientBase`.

prime example of this is Java's hashed dictionaries, e.g., `HashMap` and `HashTable`. Serialized instances of these types are inherently required to recalculate the hash values of their keys since the hashing implementation depends on native code on the operating system a Java Virtual Machine (JVM) is running on. Consider Listing 2, where from `readObject()` (line 2), keys and values are read from the stream (lines 7 and 8), and then the keys' hashes are calculated before placing them in the table (line 12). The resulting call to an arbitrary `Object.hashCode()` is no reason for concern in itself. However, it unlocks numerous additional gadgets to integrate into the chain.

```
1  class HashTable<K,V> implements Map<K,V>, Serializable {
2    private void readObject(ObjectInputStream s) {
3      readHashTable(s);}
4    void readHashTable(ObjectInputStream s) {
5      int elements = s.readInt();
6      for (; elements > 0; elements--) {
7        K key = (K)s.readObject();
8        V value = (V)s.readObject();
9        reconstitutionPut(table, key, value);
10     }}
11   private void reconstitutionPut(Entry[] t, K key, V value) {
12     int hash = key.hashCode();
13     // put value/hash in table ...
14   }}
```

Listing 2: Java's `HashTable` [42] deserialization. The highlighted lines show the gadget chain to `Object.hashCode()`.

Gadgets like `Object.hashCode()`, which both open up many further polymorphic call sites and are reachable through classes in the Java Class Library (JCL), are commonly referred to as **trampoline gadgets** [51]. In *Ysoserial* 22 of 34 deserialization gadget chain payloads rely on a trampoline gadget [15, 29]. This is an important observation to make, as it shows that already making more trampoline gadgets reachable can activate currently dormant gadget chains. We demonstrate this with a motivating example.

### 2.2 Motivating Example

Consider the *DisposableBeanAdapter* [58] in the *spring-beans* dependency in Listing 3. The `Method.invoke()` (line 25) is a security-sensitive sink as it allows invoking an arbitrary method via reflection. A gadget chain `.run()` → `.destroy()` → `.invokeCustomDestroyMethod()` → `Method.invoke()` is not found by gadget chain detection tools since `run()` is not reachable for an arbitrary subtype of `Runnable`.

However, this *dormant gadget chain* can be activated by introducing a new gadget either into the *spring-beans* dependency itself or into any other dependency being used by a software project relying on *spring-beans*. Listing 4 shows a gadget deliberately crafted for

```
1  package org.springframework.beans.factory.support;
2  class DisposableBeanAdapter implements Runnable, Serializable {
3    private final Object bean;
4    private final String beanName;
5    private String [] destroyMethodNames;
6
7    @Override
8    public void run() { destroy(); }
9
10   public void destroy() {
11     // irrelevant lines omitted
12     else if (this.destroyMethodNames != null) {
13       for (String destroyMethodName : this.destroyMethodNames){
14         Method destroyMethod =
15           determineDestroyMethod(destroyMethodName);
16         if (destroyMethod != null)
17           invokeCustomDestroyMethod(destroyMethod);
18       }
19   }}
20
21   private void invokeCustomDestroyMethod(Method destroyMethod){
22     int paramCount = destroyMethod.getParameterCount();
23     Object[] args = new Object[paramCount];
24     ReflectionUtils.makeAccessible(destroyMethod);
25     Object returnValue = destroyMethod.invoke(this.bean, args);
26     // remaining lines omitted
27   }
28  }
```

Listing 3: Dormant deserialization gadget chain in *spring-beans'* `DisposableBeanAdapter` [58]. Given the reachability of `Runnable.run()` through any serializable class on the classpath, the gadget chain can be activated.

this cause. With `hashCode()` being a trampoline gadget, the call to `Runnable.run()` at line 7 can be invoked through deserialization. An attacker could assign a `DisposableBeanAdapter` (Listing 3) to the `hashCodeGen` field of the gadget in Listing 4 and serialize it to create a malicious payload. Note how the gadget is further concealed since the `Runnable` property `hashCodeGen` is declared as `final`, thus suggesting the field cannot be arbitrarily assigned. Yet, Java deserialization bypasses default constructors and thereby allows setting final fields during deserialization [18].

```
1  class HypocriteSerializable implements Serializable {
2    private int hashCode = 0;
3    private final Runnable hashCodeGen = new Runnable() {
4      public void run() { /*...*/ }
5    };
6    public int hashCode() {
7      if (this.hashCode == 0) hashCodeGen.run();
8      return this.hashCode;
9  }}
```

Listing 4: Gadget making `Runnable.run()` reachable.

This example gives evidence of why dormant Java deserialization gadget chains make for a lucrative supply chain attack target. An attacker has ample opportunities to hide code that activates the chain. Moreover, the code additions are in a class or dependency unrelated to the vulnerable gadget that is being targeted.

### 2.3 Software Supply Chain Threats

The software supply chain comprises all dependencies, tools, and deployment infrastructure to build and deliver a piece of software. High-profile incidents like *SolarWinds* [46], *Log4Shell* [14] or *XZ Utils* [38] showed how such supply chain components can compromise many downstream targets. Specifically, when looking at the two recent cases, *Log4Shell* and *XZ Utils*, it highlights (1) widely used dependencies as the weak link in the supply chain and (2) that a compromise may occur accidentally or deliberately. In the same way, an exploitable deserialization gadget could be introduced with or without intention. Either way, the closer a dormant gadget chain is to becoming a real chain, the more likely a code change activating it is going to be unnoticed.

A previous work [68] assessed the feasibility of adding small code changes to turn immature into real vulnerabilities. While we distance ourselves from their research[3], we acknowledge that its concept is instructive to our work. In [68], the authors searched the *Linux* kernel for immature *Use-After-Free* (UAF) vulnerabilities and showed how to activate those through stealthy code changes. By analogy, we search for incomplete (dormant) gadget chains (as in Listing 3) and evaluate the effort to turn those into a full deserialization gadget chain. Of course, these changes should also be stealthy. Java's `Serializable` API has been criticized as being opaque [18, 19], which has some potential to hide gadgets. Further, one viable way of creating more gadgets is to introduce the `Serializable` interface to existing classes, thus turning them into gadgets [55]. For this reason, we investigate how the usage of Java's `Serializable` evolves in dependencies over time.

## 3 Evolution of Java's Serializable Interface

We describe our analysis pipeline towards a representative and comprehensive view for Java's *Serializable* API usage in dependencies over time and versions. To this end, we build a two-stage pipeline (see Figure 1) to first obtain a clean mapping of dependency download URLs, versions, and release dates. Then, we forward the URLs by dependency to static analysis to both analyze *Serializable* usage over time and versions as well as the reasons for these changes. The first stage is implemented in *Python* in 550 lines of code and the second stage in Java on top of *SootUp* [25] in 201 lines of code.

### 3.1 Data Collection and Cleaning

*3.1.1 Data Source Definition.* The naive approach to collecting a dataset of Java dependencies would be to crawl the *Maven Central* repository root. However, this practice leads to an unrepresentative dataset and is also wasteful:

- The repository listing gives no indication of dependency usage.
- It is hard to filter out dependencies that are unlikely to be included during runtime (e.g., test and build dependencies).
- Dependencies can get relocated within the repository listing. For instance, *servlet-api* is moved to *javax.servlet-api* and later *jakarta.servlet-api*. Such relations are crucial to produce a coherent dependency evolution history. This information is also not stored in the now six-year-old *Maven Dependency Graph* by research [3].
- The *Maven Central* repository alone comprises 51.9 TB. Taking into account all repositories listed through *Maven*, it is 202 TB.

Instead, we start from the *mvnrepository.com* web interface, which provides usage statistics, relocation information, and a dependency categorization. To our knowledge, there is no convenient API to retrieve this data, so we must resort to scraping it from the user-oriented web front. As such, we compromise with two further challenges: request rate limiting and an incomplete query mechanism for dependency popularity. The latter is caused by the /popular endpoint (see Figure 1) only paginating to page 20. Given that each page contains 10 dependency entries, this would limit us to only the 200 most used dependencies. Therefore, we additionally query all category based listings, setting a minimum usage of 100 as a threshold to filter out less popular dependencies[4].

---
[3]The publication was retracted due to unapproved human-subject research [22].

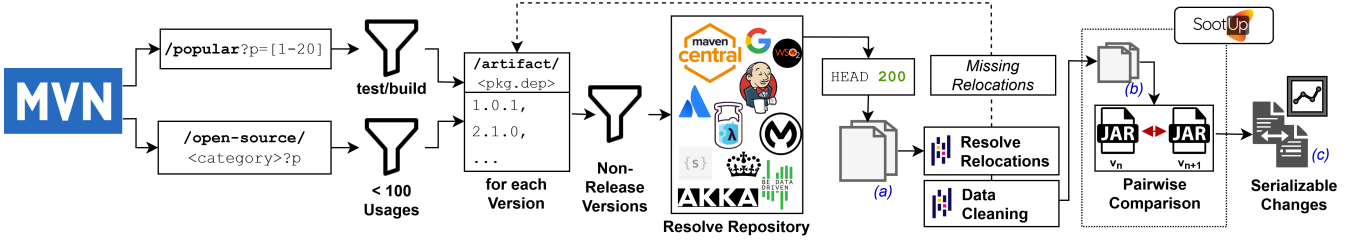

Figure 1: Serialization evolution dependency download and analysis pipeline. The blue letters (a-c) refer to intermediary artifacts: (a) all downloadable dependencies, (b) cleaned and remapped relocations, and (c) enriched with serializable changes.

*3.1.2 Web Scraping.* We preliminarily filter version tags indicating non-release versions (e.g., *beta*, *alpha*, *rc*). Note that during data cleaning we will more thoroughly remove such versions. At this stage, however, we only want to reduce the number of web requests. Further, we cannot retrieve the exact download URL by visiting all the version links on *mvnrepository.com* due to the website's request rate limiting. Instead, we construct a download URL (see Appendix A) where the repository is one of eleven repository base URLs. Most artifacts are hosted on the *Maven Central* repository. The remaining ten repositories are fallbacks for certain categories. For instance, the *Google Maven* repository hosts many *Android* dependencies that are not stored on *Maven Central*. For each dependency (-version), we probe whether a dependency is hosted with a HEAD request starting from *Maven Central* and then iterating through the fallback repositories. We also try to download the dependency in *AAR* (*Android Archive*) format if no JAR is available.

Some listings refer to the same dependency, which is indicated by the *relocated* tag. We can extract this information from scraping *mvnrepository.com* and add a reference column to point to the new location of the dependency. This allows us to merge instances such as *java-servlet-api*, *javax.servlet-api*, and *jakarta.servlet-api* to a single coherent dependency. Unfortunately, the *relocation* tag is a single-linked list. If only the newer dependencies are contained in the popularity listing, the older versions cannot be retrieved.

Through the web scraping, we retrieve 1 435 unique dependency names with 119 202 version entries (see Figure 1 (a)). From here, we identify dependencies that reference a dependency relocation, but this dependency does not exist in our dataset. We resolve the missing 40 dependency relocations, adding another 6 428 versions to the dataset. Thus, the pre-data-cleaning dataset comprises 1 475 unique dependency names with 125 630 versions.

---
[4]The reason we cannot solely rely on the category-based listings is because a select few popular dependencies (e.g., *org.renjin.stats*) are not categorized at all.

*3.1.3 Data Cleaning.* We recursively map dependency relocations to the latest dependency relocation's name. This way, we can query the complete version history across relocations. Then, since our aim is to analyze Java's *Serializable* usage over version and time, we must rigorously clean the version identifiers into a sortable scheme. Ideally, this scheme would adhere to semantic versioning [50]. However, some dependencies ignore semantic versioning entirely. For instance, *org.json* strictly uses a date format (e.g., 20250107) as a version number, which, for our purposes, is fine since it ensures correct version-based sorting. If this scheme is used only occasionally, as in the *Apache Commons Collections*, then we remove these version entries (see Table 1, row 4). After mapping relocations and version identifier cleaning, the dataset contains 1 100 unique dependencies and 111 275 version entries (see Figure 1, (b)).

| Rule | Example |
| --- | --- |
| *Release Tags* | 5.2.25.FINAL → 5.2.25 |
| *Alpha Tags* | 1.6.0-dev01 → (remove) |
| *Char Separators* | 1.5R4 → 1.5.4 |
| *Date Versions ≤ 20%* | [3.2.2, 3.2.1, …, 20040616] → [3.2.2, 3.2.1, …] |
| *Undefined Postfix* | 2686.v7c37e0578401 → 2686 |

Table 1: Version identifier cleaning rules.

For each dependency, we iterate over all versions in the dataset and load these pairwise into two *SootUp* views. Tribute to the version name cleaning enabling correct sorting; we can not only collect the number of serializable classes in the view but also determine changes to the previous version. Specifically, we determine the cause of an in/decrease of Serializable classes. The three causes are depicted in Figure 2. Going from the current version $v_n$, its successor $v_{n+1}$ could have (1) added a new serializable class or (2) implemented the Serializable interface to an existing class, which could (3) indirectly render child classes serializable. A decrease in serializable classes occurs through the inversion of the three scenarios. We enrich the cleaned dataset with the number of such change events in each dependency version.

## 3.2 Measuring Serialization Usage

In the following, we measure the trend for Serializable usage in Java dependencies over their evolution. First, we remove dependencies for which, over the entire version evolution, not a single class was serializable. This halves the dependencies to 533. For statistical analysis, we further trim dependencies that have no release in 2024 or 2025. This is especially important for evaluating the correlation of serializable class count over dependency evolution. A correlation coefficient for a dependency with only releases up to, e.g., 2016 would pollute the current trend. Hence, the dependency dataset relevant to analyzing Serializable evolution comprises 352 dependencies (*32%* of the original 1 100).

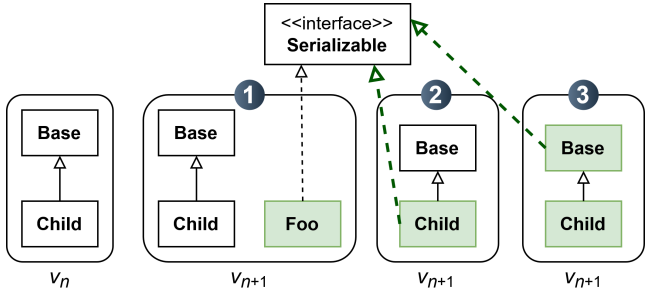

Figure 2: Causes (1-3, change in green) for an increase in serializable classes.

In Table 2, we investigate serialization evolution over time and dependency version separately. That is because version numbers are not necessarily congruent with the release date. For instance, the *Spring* framework will create a new major release but continue to release new minor releases for older major release versions. As an example, *spring-core v6.0.0* was released in November 2022, and thus earlier than *v5.3.39* (August '24) and *v5.2.25* (July '23).

| $\rho \in$ | [0.5, 1] | [0, 0.5] | [0, −0.5] | [−0.5, −1] | n.a. |
|---|---|---|---|---|---|
| **Date** | 174 (*49.43%*) | 33 (*9.38%*) | 39 (*11.08%*) | 37 (*10.51%*) | 71 |
| **Version** | 182 (*51.70%*) | 28 (*7.95%*) | 34 (*9.66%*) | 39 (*11.08%*) | 71 |

Table 2: Correlation of class serializability with dependency evolution over time/version. For 71 of 352 dependencies the standard deviation is 0, i.e., no correlation can be calculated.

Table 2 shows that for dependencies which use Java's Serializable interface and have recent releases, the majority has a strong *Pearson*-correlation for Serializable class count over time (*49.43%*) and version (*51.70%*). This set contains prominent dependencies such as *Google*'s *Guava* core Java libraries (Figure 3a), *Apache Dubbo* (Figure 3b), or *Spring Core* (Figure 3d). Yet, even for dependencies with a weak or negative correlation, serializable classes can fluctuate to a degree that makes it feasible to introduce new gadgets (see Figures 3e to 3j).

We also calculate the overall correlation of Java's Serializable usage across dependencies. To do so, we normalize the amount of data points by year, removing dependencies that do not contain at least one dependency release for every year from 2015 to 2024. 133 dependencies satisfy this condition. For dependencies with multiple releases within one year, we select a random sample. This step is mandatory to avoid skewing the overall correlation towards dependencies with many releases (e.g., the *aws-core* dependency contains 3 407 releases). Due to random sampling, we calculate the overall correlation 100 times and take the mean of this value. As such, we find a weak positive correlation of $\rho_{all} = 0.1925$.

In addition to the number of serializable classes in each dependency version, we also track the causes for an increase/decrease in serializable classes. We categorize the changes in Table 3. The addition or removal of a serializable class is by far the most common type of change (*86.83%*). Still, we find *9.36%* of change events to be related to implementing the Serializable interface – directly or indirectly (as shown in Figure 2). These changes occur in 190 or 108 dependencies, respectively. These are also the dependencies where a change in serializability is most likely to remain unnoticed.

| | **Change Events** (242 919) | | **Dependencies** (352) | |
|---|---|---|---|---|
| | *Add* | *Remove* | *Add* | *Remove* |
| **Class** | 124 424 (*51.22%*) | 86 497 (*35.61%*) | 275 | 237 |
| **Direct** | 14 640 (*6.01%*) | 6 367 (*2.62%*) | 190 | 109 |
| **Indirect** | 8 090 (*3.33%*) | 2 901 (*1.19%*) | 108 | 74 |

Table 3: Changes responsible for increase/decrease of serializable class count in dependencies with releases in 2024/25.

## 3.3 Key Observations

We analyzed the usage of Java's Serializable interface over time and version in a dataset of 1 100 widely used dependencies. It leads us to (a) formulating a **key challenge to a comprehensive view** on deserialization gadget chain research and (b) implications for gadget-chain-based **supply chain attacks**. To do so, we first define four dependency datasets ($D \subset C \subset B \subset A$):

- *A* **Gadget Providers** (533) – dependencies containing at least one serializable class.
- *B* **Active Gadget Providers** (352) – *Gadget Providers* with a recent release in 2024/25.
- *C* **Volatile Gadget Providers** (283) – *Active Gadget Providers* with a fluctuating amount of serializable classes over the dependency evolution (due to all causes outlined in Figure 2).
- *D* **Volatile Stealthy Gadget Providers** (208) – *Volatile Gadget Providers* with at least one change event related to implementing/removing the Serializable interface to/from a class.

To increase visibility of deserialization gadget chains in Java, we must be able to search for gadget chains in the entirety of set *A* across dependencies and also across versions if dependencies are contained in set *C*. Since we continue to focus on *dormant gadget chains*, solving this challenge is out of scope for this work. However, we provide the datasets to the research community as an open research challenge (see Section 8). Although such an analysis does not eliminate gadget chains within downstream software projects, it solidifies the supply chain by providing visibility to gadget chains caused by dependencies. Also, with it, *Ysoserial* [15] can be brought to an up-to-date version.

In this work, focusing on gadget-chain-based supply chain attacks, we continue to work on dependencies in datasets *A*, *B*, and *D* individually. These are the datasets that directly map to our three gadget injection patterns. Dataset *C* is irrelevant because it would only apply to a pattern for adding new serializable classes to a target dependency. However, it is hard to formulate a universal pattern for the addition of a functionally independent class.

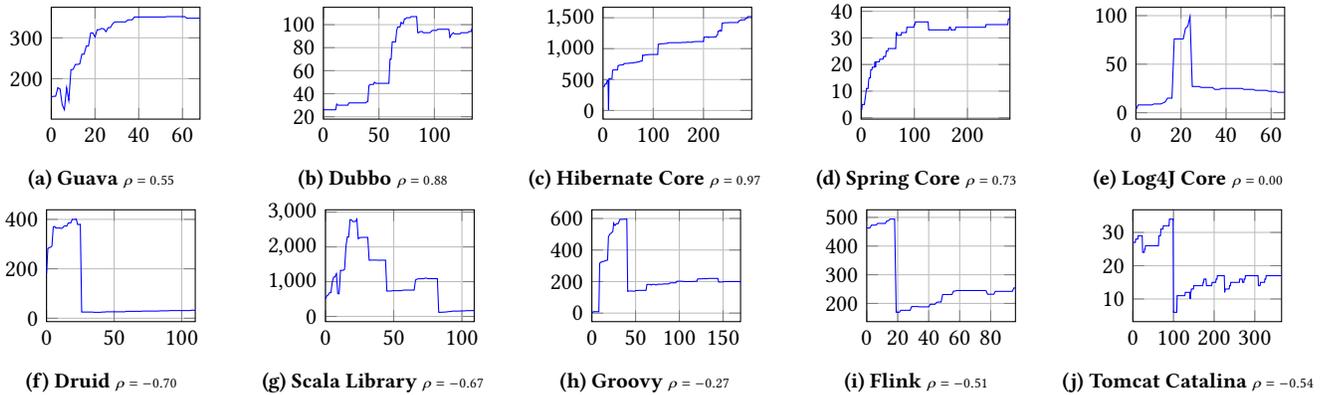

Figure 3: Visualization examples for serializable class count over dependency version. The x-axis represents the version index; the y-axis the number of serializable classes.

## 4 Activating Dormant Gadget Chains

We aim to activate currently dormant deserialization gadget chains through three modification patterns. Our approach is inspired by *FixReverter* [71] and *Hypocrite Commits* [68]. However, this work is, to the best of our knowledge, the first bug injection research dedicated to Java deserialization gadget chains.

Figure 4 outlines our process for gadget injection and evaluation of success rate. For each dependency in the dataset, we detect injection sites and apply the three patterns (1-3, further detailed in Section 4.1). We then run gadget chain detectors on the original and modified JAR files which results in two possibly distinct outputs. The delta of detections between these outputs is a soft proof of the injection patterns' success.

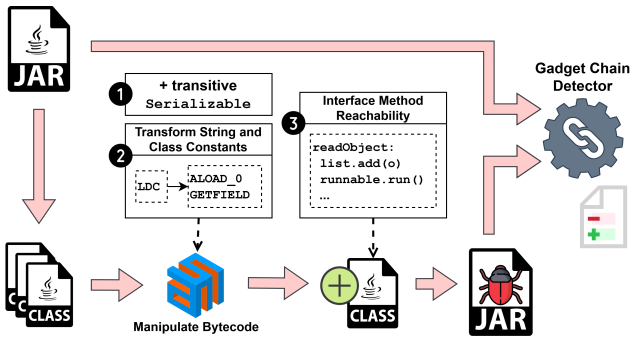

Figure 4: Deserialization Gadget Injection Framework.

From there, we provide hard evidence by manually analyzing 1 990 detected (dormant) gadget chains in 126 dependencies (Section 4.4). In Section 4.5, we illustrate the interplay of our three injection patterns with one of the true positives we identified as a case study.

### 4.1 Injection Patterns

*4.1.1 Transitive Serializability.* While direct changes to class serializability may occur over a dependency's evolution, from an attacker's point of view, making a superclass or interface serializable is a more desirable, hidden approach. A good example of this is the *Apache Commons Collections*' `InvokerTransformer` and `InstantiateTransformer`. As the notorious enablers of seven *Ysoserial* gadget chains, their serializability was removed in one patch[5], leaving a warning comment to never make these classes serializable again. Under these circumstances, re-implementing the *Serializable* interface in the two classes is unlikely. However, what if the `Transformer` interface were to become serializable? Given the history of *Commons Collections* gadget chains, such a patch being accepted may still be contrived – albeit, much less than adding the *Serializable* interface directly. Now, let us consider the dependency set of *volatile stealthy gadget providers* (Section 3.3, *D*), which have had transitive serializability changes in the past and have not been subject to gadget chain detection. A maintainer might consider the consequences of serializability to the applied class/interface itself but miss the implications to inheriting components. An attacker can leverage this to conceal their target.

We apply the *transitive serializability* pattern by implementing the `Serializable` interface to all abstract classes and interfaces in a dependency.

*4.1.2 Final Properties.* A property declared as `final` may only be assigned once [43]. Taking Listing 5 as an example, one could assume the classes `NotVuln` and `Vuln` to be semantically equivalent. Yet, Java deserialization bypasses the default constructor [18], which otherwise would set `methodClass` and `methodName` (lines 7 and 8) to the given constants. This allows arbitrarily setting both `final` properties, leading to arbitrary method invocation at line 11. Note that `methodName` should not be defined as a simple string literal because, if final, it is considered a compile-time constant that cannot be modified [43]. A maintainer unfamiliar with the potential of setting `final` fields through deserialization could easily miss a malicious refactoring as in Listing 5.

For gadget injection, an attacker can specifically target `String` and `Class` constants used in methods and transform those into properties. We replace the `LDC` bytecode instruction for loading from the constant pool with an equivalent ALOAD 0 (0 refers to `this`) and

---

[5]https://github.com/apache/commons-collections/commit/e585cd0433ae4cfbc56e58572b9869bd0c86b611#diff-2d13b1592fb865090f134fe9d88dee2cb2e24170a5338d5df79a495b34b207a9

```java
1  class NotVuln implements Serializable{
2    public void method(Object o) {
3      Method method = OtherClass.class.getMethod("m");
4      method.invoke(o);
5  }}
6  class Vuln implements Serializable{
7    private final Class methodClass = OtherClass.class;
8    private final String methodName = new String("m");
9    public void method(Object o) {
10     Method method = methodClass.getMethod(methodName);
11     method.invoke(o);
12 }}
```

**Listing 5: Extracting constant pool values into `private final` properties.**

GETFIELD, which refers to the newly added final field [44]. The modification essentially allows an attacker to control constant values, opening up a broad range of security-sensitive sink methods requiring a tainted *String* or *Class*. Therefore, while we primarily target common usage patterns of the Java reflection API [33, 66], we also cover *String* constants used in, e.g., Runtime.exec(<String>) or custom *ClassLoaders*. Any dependency that is active in development (i.e., *B - Active Gadget Providers*) is an injection target.

*4.1.3 Interface Method Reachability.* Of all three change patterns, this modification is the stealthiest. The aim is to introduce new trampoline gadgets into software so that a multitude of further gadgets become reachable. The change is difficult for maintainers to detect, especially since the caller of the new trampoline gadget could be in a different dependency. Consider a subtype of `Iterator` with an implementation of `Iterator.next()` that leads to a gadget chain. An attacker (or a developer unintentionally) could create a serializable class with a generic `Iterator` property. As long as this class coexists with the dependency containing the `Iterator`-gadget on an application's classpath, this application now contains a full deserialization gadget chain. For this reason, this pattern is applicable to all *Gadget Providers* (dataset *A*).

For each serializable class in the dependency, we extract all implemented interfaces. Then, we add all interface methods related to the Java SDK API to a caller gadget (e.g., Listing 6). For an attacker, this ensures that no other dependencies must be present in the dependency they use as an injection target. We consider interfaces to be part of the Java SDK API if they reside in the java.* or javax.* packages (similar to [35]). Note that we add all interface methods even if a serializable class does not explicitly implement those due to inheritance scenarios as in Listing 7. Also, the `Caller` gadget contains a constructor to set all properties as well as `hashCode` as a trampoline deserialization entry point. This is specifically necessary for one of the tools we use during experimentation – *Crystallizer* [60] as its static analyzer uses a trampoline heuristic, and the fuzzer only sets properties via available constructors [28]. Finally, we repackage the `Caller` gadget into the original dependency JAR file.

### 4.2 Experimentation

*4.2.1 Gadget Injection Framework.* We implement the gadget injection in Java using *ASM* [4] and *SootUp* [25] (736 LoC). With this setup, we can apply modifications to bytecode-compiled dependencies. For the *transitive Serializable* pattern, we detect `abstract` and `interface` classes in a dependency with *SootUp* and then use a

```java
1  class Caller implements Serializable {
2    public Iterator iterator; public Runnable runnable;
3    public Object object; public Function function;
4    // additional interface properties ...
5    public Caller(Iterator _iterator, Runnable _runnable, ... ) {
6      this.iterator = _iterator;
7      this.runnable = _runnable;
8      // ...
9    }
10   @Override
11   public int hashCode() {
12     iterator.next();
13     function.apply(object);
14     // additional interface method calls ...
15 }}
```

**Listing 6: Arbitrary JCL interface method reachability gadget.**

```java
1  class Base implements java.lang.Runnable {
2    @Override
3    public void run() { /*...*/ }
4  }
5  class Child extends Base implements Serializable {}
```

**Listing 7: The method `Base.run()` is a gadget despite `Base` not being serializable as it is inherited to `Child.run()`.**

simple *ASM* class visitor to append the `Serializable` interface. Swapping constants (pattern 2) requires a combination of a class and method visitor to first add property fields and then, in the methods themselves, replace LDC instructions to retrieve the original constant pool value from the newly defined fields. Finally, for the *Interface Method Reachability* pattern, we implement a helper Java SootUp program to identify implemented interfaces of serializable classes and generate the source code of the `Caller` gadget. We then compile the source code and repackage it with the original dependency JAR. Table 4 provides the minimum, maximum, average amount, and standard deviation of modifications per dependency. Observe that for pattern (3), which is applicable to all datasets, for the larger datasets *B* and *A*, the average amount of interfaces added to the `Caller` gadget is declining. This indicates how dataset *D* (**Volatile Stealthy Gadget Providers**) is the most lucrative target for injection, regardless of the pattern.

| Pattern and Dataset | | min | max | $\bar{x}$ | $\sigma$ |
|---|---|---|---|---|---|
| **(1) - classes modified** | D | 1 | 5 139 | 233.84 | 483.18 |
| **(2) - classes modified** | B | 0 | 11 704 | 308.51 | 832.72 |
|  | D | 2 | 11 704 | 414.86 | 1 038.77 |
| **(3) - interfaces in `Caller`** | A | 0 | 42 | 2.47 | 5.73 |
|  | B | 0 | 42 | 2.62 | 6.02 |
|  | D | 0 | 42 | 3.38 | 6.65 |
| **all - interfaces in `Caller`** | B | 0 | 88 | 5.10 | 9.97 |
|  | D | 0 | 88 | 7.69 | 11.98 |

**Table 4: Statistics for applying patterns (1) - (3)**

We also build a dependency containing all modifications. Pattern (1) leads to more classes becoming serializable, which can increase the number of interfaces added by the `Caller` gadget in pattern (3). Therefore, we proceed in two stages, creating an intermediary JAR with patterns (1) and (2) applied, followed by the pertaining analysis and gadget construction of (3). As seen in Table 4 (last row), this more than doubles the average number of interfaces being called from the Caller gadget.

*4.2.2 Gadget Chain Detection Tools.* We run the three state-of-the-art gadget chain detectors, *Tabby* [10], *Crystallizer* [60], and *AndroChain* [29], on the untouched and modified dependencies. Thereby, we aim to identify whether the tools report surplus gadget chains with the injection patterns applied. We specifically choose these three tools due to their availability, recency, and diverse methodology. We exclude the state-of-the-art tool *JDD* [8] from our study due to its fuzzer module not being available [9]. Instead, we take *Crystallizer* as the next-recent dynamic gadget chain detector. The tool employs a fuzzer to prune false positive gadget chains from the initial static analysis, which makes it the most precise of the three tools in our study. Conversely, *AndroChain* puts a strong emphasis on soundness. It is not constrained by a maximum gadget chain depth, which heavily affects *Crystallizer* [28], and partially *Tabby* [10]. *Tabby* is the currently most popular gadget chain detector (1 400 *GitHub* star rating [65]). This is likely due to the high degree of reusability from its *Deserialization-Aware Call Graph* (DA-CG) in a *Neo4J* database. In fact, two further closed-source research projects leveraged *Tabby* to improve gadget chain detection [6, 7].

*AndroChain* is provided as a standalone executable JAR file. Thus, the analysis workload can be efficiently distributed with *GNU Parallel* [61] onto the 128 CPUs of the experimentation server[6]. Since both *Crystallizer* and *Tabby*'s *Neo4J* database are docker-contained, we can parallelize by launching multiple containers and orchestrating the dependency analysis tasks to those. Moreover, *Crystallizer* is recommended to run 25 hours on each dependency – one hour for dynamic sink identification and 24 hours for fuzzing gadget chains. We can speed up the analysis by first running *Crystallizer* only for one hour in dynamic sink analysis and static gadget chain detection. If this part yields no results[7], there is no reason to run the full 24-hour fuzzing campaign. For *Tabby*, we configure the toolchain described by the maintainers: *tabby* [65] for generating the DA-CG, *tabby-path-finder* [63] as the custom *Neo4J* procedure plugin, and *tabby-vul-finder* [64] to load the DA-CG and execute queries. Once a cluster of *Neo4J* containers is launched with the *path-finder* plugin, we again leverage *GNU Parallel* for DA-CG generation, database import, and querying.

The tool instrumentation, container orchestration, and result parsing is implemented in 590 lines of *Python* code.

### 4.3 Results

Table 5 shows the results of running *Tabby*, *Crystallizer*, and *AndroChain* on the three datasets and with the individual modifications applied. It represents how many of the modified dependencies contained additional gadget chains in comparison to the unmodified dependency, according to the respective gadget chain detectors. These additional detections are a soft proof for dormant gadget chains. Even though it does not indicate whether these are true positives, it shows that further execution paths were considered by the tools during analysis. Overall, the tools found potential dormant gadget chains in 139 (*26.08%*) dependencies for dataset *A*, 114 (*32.39%*) for dataset *B*, and 99 for dataset *D* (*47.60%*).

*Crystallizer*'s results are underwhelming for a few reasons. Recall that *Crystallizer* starts with a dynamic identification of potential sink methods. This preliminary fuzzing stage shrunk the number of eligible dependencies to 92 (*17.3%*) in dataset *A*, 67 (*19.0%*) in *B*, and 49 in *D* (*23.6%*). Further, *Crystallizer* detects chains only up to a length of five gadgets [28, 59]. The gadget chain fuzzer uses constructors and setter methods to set properties [8]. While we specifically accommodated for this constraint in our modification patterns (see Section 4.1.3), *Crystallizer* still fails to set all possible properties (i.e., using Java Reflection) in the gadgets contained by the dependency as is. Also note that due to the non-determinism of fuzzing, *Crystallizer* reported alleged additional gadget chains in two more dependencies when applying only modification (1) as opposed to applying all modifications (Table 5, row D). For all the above-mentioned reasons, we do not consider *Crystallizer*'s results representative for the success rate of our injection patterns. We further prove this by manual verification of *Tabby*'s and *AndroChain*'s gadget chains in Sections 4.4 and 4.5.

Considering *Tabby*'s and *AndroChain*'s results, all three modification patterns appear to be a viable strategy to activate dormant gadget chains. *AndroChain* may have a small delta to the detections in the unmodified dependencies for pattern (2) due to it not filtering outputs based on taint[9]. Here, it proves crucial to have the comparison to a different state-of-the-art tool, i.e., *Tabby*. For *Tabby*, all patterns have approximately similar viability. One should not be deceived by the declining relative amount of dormant gadget chain detections in the larger datasets *A* and *B* with *all* modifications applied. This is likely due to not applying patterns (1) and (2). However, when regarding pattern (3) in isolation, there is an indication of diminishing returns for the larger datasets.

### 4.4 Verifying Gadget Chain Detections

Table 5 provides a soft proof of the individual patterns' success rates. This is backed by 830 new gadget chain detections by *Tabby*, 868 by *Crystallizer*, and 2 785 by *AndroChain*. In this section, we provide hard evidence for our gadget injection framework by manually determining true positives from the tools' results.

We start by filtering out gadget chains, leading to less promising sink methods. For instance, *Tabby* and *AndroChain* consider Class.forName() and Class.getMethod() to be sinks. That is likely because they may be succeeded by a call to Method.invoke() – a typical usage pattern of Java Reflection [28, 33]. Yet, in isolation, these methods have no strong security implication [15]. We also disregard sink methods related to *InputStreams*. Verifying gadget chains with these sinks is more time-consuming since it requires not only verifying the gadget chains' control flow but also the subsequent usage of the *InputStream* in the program. Overall, we restrict ourselves to 23 interesting sink methods (complete listing, see Appendix B).

---

[6] *Debian Linux* 5.10.218 OS with a 64-core AMD EPYC 7713P (2.00GHz) processor and 995GB RAM
[7] I.e., the intermediary results/concretization/gadgetDB.store file is missing.
[8] [59] – */src/dynamic/Meta.java*, line 850ff. and line 1088ff.
[9] The tool still outputs the propagated taint value in its results, which we could use to calculate a delta [29].

| | Tabby [10] | | | | Crystallizer [60] | | | | AndroChain [29] | | | |
|---|---|---|---|---|---|---|---|---|---|---|---|---|
| | (1) | (2) | (3) | all | (1) | (2) | (3) | all | (1) | (2) | (3) | all |
| A (533) | - | - | 45 (*8.4%*) | 56 (*10.5%*) | - | - | 0 (*0.0%*) | 8 (*1.5%*) | - | - | 48 (*9.0%*) | 101 (*18.9%*) |
| B (352) | - | 42 (*11.9%*) | 33 (*9.4%*) | 44 (*12.5%*) | - | 7 (*2.0%*) | 0 (*0.0%*) | 8 (*2.3%*) | - | 9 (*2.6%*) | 29 (*11.1%*) | 82 (*23.3%*) |
| D (208) | 33 (*15.9%*) | 32 (*15.4%*) | 27 (*13.0%*) | 35 (*16.8%*) | 8 (*3.8%*) | 4 (*1.9%*) | 0 (*0.0%*) | 6 (*2.9%*) | 66 (*31.7%*) | 7 (*3.4%*) | 23 (*11.1%*) | 74 (*35.6%*) |

**Table 5: Dependencies with additional gadget chain detections in the datasets $A$ (Gadget Providers), $B$ (Active Gadget Providers), and $D$ (Volatile Stealthy Gadget Providers), when applying the changes (1) - (3) in isolation, or *all* applicable changes combined.**

This leaves us with 1 990 gadget chains in 126 dependencies after filtering. We distribute the workload of verifying these gadget chains on three researchers in our team. Each researcher receives a set of 42 dependencies with the respective tools' gadget chain detections. Upon confirming one true positive, we skip the remaining detected gadget chains for this dependency. This is because we are interested in a measure of how many dependencies are susceptible to our injection patterns. The manual analysis involves decompilers (*JADX* [57]) and debugging tools (*IntelliJ* [24]) for PoC generation. The latter is reserved for complex cases (e.g., Appendix B, Table 15) where we cannot be certain from static analysis whether the conditions for a certain control flow can be met. We cannot do this in every case since creating a PoC for a single gadget chain takes around two hours. Overall, our analysis spans over a work week.

Our combined efforts yield verified gadget chains in 53 dependencies. We list the gadget chains in Appendix C. Figure 5 shows how many of these verified dormant gadget chains could be activated by applying the change patterns individually or in combination. Indeed, each pattern is viable on its own for at least one dependency. More specifically, 26 (*49.06%*) of the dormant gadget chains require only one of the three change patterns. It shows that for half of the true positive cases, dependencies are on the verge of containing a security-critical gadget chain. The most common missing prerequisite is *interface method reachability*, which, as we discussed in Section 4.1, is the easiest to satisfy from an attacker's perspective.

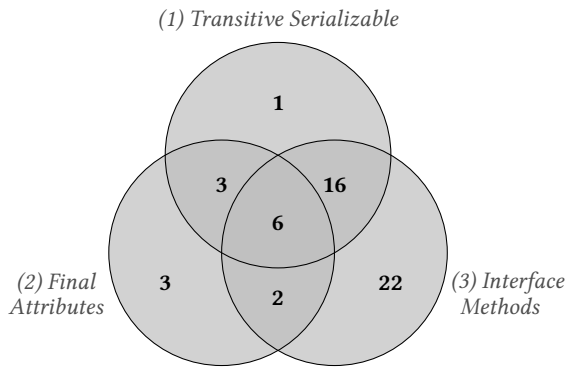

**Figure 5: Venn diagram of how many dependencies were activated with which modification(s).**

Table 6 zooms in on the dormant gadget chains, which can be activated through *interface method reachability*. Many of the listed method calls (e.g., `AutoCloseable.close()` or `Iterator.next()`) have a benign appearance. This drastically increases the odds of such a call edge being introduced. Moreover, in the cases where no further changes are required (✓ in Table 6), one can actually assume the interface methods as sink methods for a gadget chain detector. Upon finding a chain to these interface sinks, the full gadget chain can be constructed by combining it with an applicable dormant gadget chain for *interface method reachability* (see Appendix C.3).

| Interface Method | Count | Single |
|---|---|---|
| `java.lang.Runnable.run()` | 10 | ✓ |
| `java.util.concurrent.Callable.call()` | 1 | ✓ |
| `java.awt.event.ActionListener.actionPerformed()` | 2 | ✓ |
| `java.beans.PropertyChangeListener.propertyChange()` | 3 | ✓ |
| `java.lang.reflect.InvocationHandler.invoke()` | 6 | ✓ |
| `javax.sql.DataSource.getConnection()` | 3 | ✓ |
| `javax.sql.XADataSource.getXAConnection()` | 1 | ✓ |
| `java.lang.AutoCloseable.close()` | 6 | ✓ |
| `javax.sql.RowSet.rollback()` | 3 | ✓ |
| `javax.xml.transform.Transformer.newTransformer()` | 1 | ✓ |
| `java.sql.Connection.isValid()` | 1 | |
| `java.util.Iterator.hasNext()` | 4 | |
| `java.util.Iterator.next()` | 2 | |
| `java.lang.Iterable.iterator()` | 1 | |
| `java.util.Map.put()` | 1 | |
| `java.io.Flushable.flush()` | 1 | |

**Table 6: Java interface methods activating dormant gadget chains. A tick (✓) in the *single* column signifies that the activation requires no additional change patterns.**

### 4.5 Case Study - Apache OpenJPA

We exemplify how to activate a dormant gadget chain in the *Apache OpenJPA* [62] dependency. *Apache OpenJPA* is an alternate Java Persistence API used by 327 downstream projects on the *Maven Repository*. We specifically demonstrate this dormant gadget chain (see Table 7) because it incorporates all three modification patterns. While this makes the gadget chain more difficult to activate, it highlights the relevance of all three patterns in a single coherent example. For a dormant gadget chain requiring but a single modification, we refer to the motivating example in Section 2.2. In the following, we first describe the dormant gadget chain and then list the changes that were applied to activate it.

```
java.util.Iterator.hasNext()
↪ org.apache.openjpa.jdbc.meta.strats.LRSProxyMap$ResultIterator.hasNext()
  ↪ org.apache.openjpa.jdbc.sql.MergedResult.next()
    ↪ org.apache.openjpa.jdbc.sql.LogicalUnion$ResultComparator.getOrderingValue(Result, int)
      ↪ org.apache.openjpa.jdbc.sql.LogicalUnion$ResultComparator.getOrderingValue(Result, Object)
        ↪ org.apache.openjpa.jdbc.sql.PostgresDictionary.getObject()
          ↪ java.lang.reflect.Method.invoke()
```

**Table 7: Apache OpenJPA [62] dormant gadget chain.**

*4.5.1 Gadget Chain Description.* Listing 8 shows the entry point in `ResultIterator.hasNext()`. This is a perfect target for a hidden

Caller gadget calling `hasNext()` on a generic Java iterator. The gadget chain continues by invoking `Result.next()` at line 9.

```java
package org.apache.openjpa.jdbc.meta.strats;
class LRSProxyMap {
  private class ResultIterator implements Iterator, Closeable {
    private final Result[] _res;
    private Boolean _next = null;
    @Override
    public boolean hasNext() {
      if (_next == null)
        _next = (_res[0].next()) ? Boolean.TRUE : Boolean.FALSE;
      return _next;
    }
}}
```

Listing 8: `LRSProxyMap$ResultIterator.hasNext()`.

Then, using a `MergedResult`, one can divert the execution flow to a `ResultComparator` (Listing 9, line 16). Take note of how an attacker has control of the properties `_status` and `_pushedBack` to satisfy the conditional statements at lines 10 and 13.

```java
package org.apache.openjpa.jdbc.sql;
public class MergedResult implements Result {
  private final Result[] _res;
  private final byte[] _status;
  private final ResultComparator _comp;
  private boolean _pushedBack = false;

  @Override
  public boolean next() throws SQLException {
    if (_pushedBack) { ... }
    if (_comp == null) { ... }
    for (int i = 0; i < _status.length; i++) {
      switch (_status[i]) {
        case NEXT:
          if (_res[i].next())
            _order[i] = _comp.getOrderingValue(_res[i], i);
          break;
      }
    }
  }
  public interface ResultComparator extends Comparator {
    Object getOrderingValue(Result res, int idx);
  }
}
```

Listing 9: `MergedResult.next()`

In Listing 10, the gadget chain proceeds with the `ResultComparator` implementation in the `LogicalUnion` class. At line 10, it shows that the `Result` passed to the comparator in `MergedResult` needs to be of type `ResultSetResult`. So, even if it is not visible in the gadget chain (Table 7), we must ensure this class is also serializable so it can be used during payload construction. The `ResultSet` retrieved at line 10 is further passed to a call to `DBDictionary.getObject()` via lines 12 and 16.

Finally, we use a `PostgresDictionary`'s implementation of `getObject()` to achieve arbitrary method execution. In Listing 11, the string constants at lines 7 and 10 present themselves to be extracted as `final` attributes according to the second modification pattern. This bypasses both the class name constraint to the object retrieved from the `ResultSet` (line 5) and the method name constraint. As popularized by *Ysoserial* gadget chains, we can supply a `TemplatesImpl` instance to achieve arbitrary code execution from the arbitrary method invocation [15].

```java
package org.apache.openjpa.jdbc.sql;
private static class LogicalUnion.ResultComparator
  implements MergedResult.ResultComparator {

  private final List[] _orders;
  private final DBDictionary _dict;

  @Override
  public Object getOrderingValue(Result res, int idx) {
    ResultSet rs = ((ResultSetResult) res).getResultSet();
    if (_orders[idx].size() == 1)
      return getOrderingValue(rs, _orders[idx].get(0));
    ...
  }
  private Object getOrderingValue(ResultSet rs, Object i) {
    return _dict.getObject(rs, (Integer) i + 1, null);
  }
}
```

Listing 10: `LogicalUnion$ResultComparator`

```java
package org.apache.openjpa.jdbc.sql;
public class PostgresDitionary extends DBDictionary {
  @Override
  public Object getObject(ResultSet rs, int column, Map map) {
    Object obj = super.getObject(rs, column, map);
    if (obj.getClass().getName()
        .equals("org.postgresql.util.PGobject")) {
      try {
        Method m = obj.getClass()
          .getMethod("getType", (Class[]) null);
        Object type = m.invoke(obj, (Object[]) null);
      } catch (Throwable t) { ... }
    }
    return obj;
  }
}
```

Listing 11: `PostgresDitionary`

*4.5.2 Required Change Patterns.* Figure 6 visualizes the payload construction for the dormant *OpenJPA* gadget chain. Thereby, components highlighted in green refer to the locations where one of the modification patterns was applied. Overall, three interfaces were made serializable, two string literals were extracted as `final` properties in `PostgresDicitionary`, and a trampoline call edge to `Iterator.hasNext()` was introduced.

Making `LRSProxyMap$ResultIterator`, `MergedResult`, and `ResultSetResult` serializable requires only the addition of *Serializable* to *OpenJPA*'s `Closeable` since all three gadgets inherit from it. It illustrates quite well how changing the serializability of a single interface (or class) can have a substantial impact on its descendants. Additionally, we provide transitive serializability via `Configurable` and `MergedResults$ResultComparator` to `PostgresDictionary` and `LogicalUnion$ResultComparator`, respectively. The changes are not far-fetched considering that the number of serializable classes in *Apache OpenJPA* has been steadily increasing from 310 in its first release to 637 in the current version.

## 5 Discussion

### 5.1 Threats to Validity

*5.1.1 Injection Patterns.* In this work, we evaluated three specific patterns to enable gadget chains in real-world dependencies. These patterns were chosen due to their trivial injection conditions. This enabled us to apply the changes automatically on a large scale.

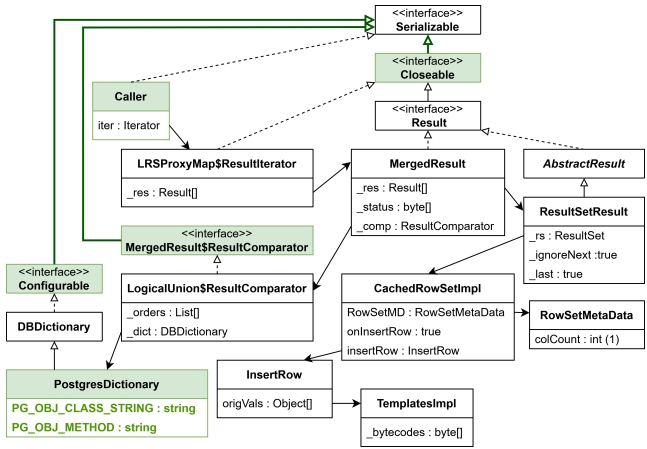

**Figure 6: Payload visualization for dormant *OpenJPA* gadget chain. Modifications are highlighted in green.**

However, we do not claim that these are the only changes that enable dormant gadget chains. A study on 19 *Ysoserial* gadget chains revealed that four change types led to the introduction of further gadget chains: adding a class or method or changing class serializability or access modifiers [55]. Through the *transitive serializable* pattern, we implemented an obfuscated class serializability change. Also, the *Caller* gadget to invoke interface methods is, in essence, the addition of a new class. There are definitely many more ways to add a class or method such that it leads to a gadget chain in the underlying dependency. Looking at the analysis in [55], it is challenging to find clear patterns for these changes. The addition of entirely new classes or methods, which also serve a functional purpose within the dependency (e.g., TransformingComparator in the *commons-collections*) requires domain-specific knowledge, thus making it quite difficult to automate. Still, for further investigation, one could consider anonymous or generated classes as in the *Clojure* gadget chain [15]. Such gadgets may only become visible in the compiled dependency and would, thereby, slip under the radar of a maintainer reviewing a malicious commit.

It is hard to empirically assess the stealthiness of our three injection patterns without compromising the integrity of open-source software development (e.g., as in [68]). However, we can deduce a hierarchy from the injection location in relation to the thereby activated gadget. Pattern (2) - *Final Properties* is applied to the gadget class itself, making it the easiest to spot during review. Then, while pattern (1) - *transitive serializability* is applied to a different class but within the same dependency, pattern (3) - *interface method reachability* has no constraints at all. By this logic, we argue that, at the very least, patterns (1) and (3) meet the requirements of stealthiness. Recall that we applied pattern (1) only to the dataset of *Volatile Stealthy Gadget Providers* (*D*), which have made changes to class serializability over their evolution. Pattern (2) is tricky to assess. It requires a developer knowing that by specification, Java final attributes are invariant [43], while being unaware that Java serialization ignores this property. According to *Goetz*, such a mistake is feasible [18], albeit our specific refactoring of *String* and *Class* literals may still be suspicious. We should still consider that an attacker would put more thought into hiding a gadget in final attributes than our pattern can simulate. Regardless, patterns (1) and (3) proved sufficient in the majority (*73.58%*, see Figure 5) of cases to activate dormant gadget chains.

*5.1.2 Dependency Dataset Size.* We analyzed popular dependencies from the *Maven* repository. This led us to define four comprehensive datasets (see Section 3.3) that we propose as a new baseline for gadget chain detection research. They encompass dependencies that, by containing serializable classes, have the potential to be leveraged towards a deserialization gadget chain. Of course, our analysis can be extended to all available dependencies and, furthermore, open-source software projects. Given that the baseline has been the 41 dependencies in *Ysoserial*, we believe it is a reasonable step to first focus on a comprehensive view of high-profile targets in the supply chain. This is exactly what we aimed for by filtering out the most popular dependencies from *Maven*.

*5.1.3 Gadget Chain Detection.* Our choice of gadget chain detectors is motivated by the *Gleipner* publication [28], which tested all, at that time, available gadget chain detection tools on a synthetic benchmark. According to their work, *Crystallizer* [60] is the most precise and *Tabby* [10] the most sound tool available. We excluded *JDD* [8] from our study since their injection object construction (IOCD) fuzzer is not available on their repository. However, we included the recent gadget chain detector, *AndroChain* [29], which detects more gadget chains than *Tabby* on *Ysoserial* as a benchmark. In doing so, we cover a reasonable proportion of the state-of-the-art in gadget chain detection. The omission of other gadget chain detectors in this work should not deter us. If anything, it would have led to even more detections of dormant gadget chains.

*5.1.4 Single Dependency Analysis.* The results of applying the three modification patterns to single dependencies prove that it is possible to activate dormant gadget chains with these changes. This was the main objective of our study. However, the attack surface could be drastically increased by incorporating cross-dependency analysis. Indeed, in *Ysoserial* 15/34 (*44.12%*), gadget chain payloads rely on more than one dependency [15]. For cross-dependency analysis, one could extract commonly grouped dependency clusters from open-source software build files and then search for gadget chains in these clusters. Moreover, regarding the *trampoline method reachability* pattern, instead of only considering interfaces in the JCL, one could also include interfaces of upstream dependencies.

## 5.2 Implications and Future Work

Our work confirms that, despite its flaws and security risks, the *Serializable API* remains a widely used part of Java-based software. The analysis in Section 3 proves that *Serializable* usage is, if anything, increasing. This appears to be a predicament for the Java platform. *JVM* providers (e.g., *Oracle*, *OpenJDK*, and *IBM*) cannot supersede the current implementation of the *Serializable API* when its usage is widespread, whereas users continue to rely on it because it is the de facto serialization standard provided by *JDK* vendors.

Our three gadget injection patterns successfully reactivated dormant gadget chains in a range of popular dependencies. This statement is supported by (1) a high estimate of 139 dependencies, which contained additional chains according to gadget chain detectors, and (2) a low estimate of 53 dependencies for which we could manually verify dormant gadget chains. While the low estimate is a solid ground truth, the high estimate gives a broader view of potential new execution paths. Either way, this proves dormant gadget chains as a hidden liability to software. Moreover, 26 (*49.06%*) of the verified dormant gadget chains required only one of our three modification patterns to be activated. These chains are on the verge of becoming a real vulnerability.

In light of these observations, we should direct our attention to mitigation techniques against deserialization-based attacks. As of now, we do not know to which extent software providers implement *ObjectInputFilter*s [52, 53]. For *Android*, it shows that app developers have not fully caught up with the *Android-SDK*-provided type-safe deserialization methods [20, 29]. We also emphasize the implications of our *Interface Method Reachability* pattern. In our view, it is almost trivial for an attacker to hide a gadget calling a new trampoline method (e.g., Iterator.hasNext()) in another dependency. Maintainers have no control over such changes. Moreover, such a gadget may actually already exist in a dependency – something we aim to search for in future work. Either way, under these conditions, we believe a dormant gadget chain that can be activated with this pattern necessitates remediation.

## 6 Related Work
### 6.1 Java Deserialization Gadget Chains

A wide range of gadget chain detection tools have been developed with diverse methodologies. The main distinction between the approaches is whether they purely rely on static analysis [2, 5, 7, 10, 21, 29, 32, 34, 36, 37, 54, 67] or additionally employ a fuzzer for validation [6, 8, 51, 60]. The tools are almost exclusively used on the dependencies in *Ysoserial*, which narrows the view on Java deserialization gadget chains to an outdated subset of the entire dependency landscape [28, 29]. The large-scale analysis of gadget chains in the *Android Open Source Project* [29] stands as an exception. While the authors did not find any exploitable gadget chains, they observed two examples of dormant gadget chains in the *RxJava* dependency. On this basis, they propose investigating the viability of deserialization gadget chains as a supply chain attack.

By analyzing the evolution of Java's *Serializable* interface usage and deducing the relevant dependencies for gadget chain detection, we set a target towards a holistic view (Section 3). It is thereby also an extension to [27, 55], which show that the presence of deserialization gadget chains is volatile over a dependency's evolution.

### 6.2 Supply Chain Attacks
There is ample evidence for the increasing risk of software supply chain attacks [14, 38, 46]. The *Backstabber's Knife Collection* [41] entails an analysis of 174 malicious packages on package repositories. Together with the work of *Ladisa et al.* [30] it has created a solid understanding of attack types and targets in a software supply chain. Specifically, our injection patterns for Java deserialization gadget chains fall into the categories *AV-100 Develop and Advertise distinct Malicious Package from Scratch* and *AV-304 Make immature Vulnerability Exploitable* [30].

The concept of making immature vulnerabilities exploitable directly relates to [68] and is adjacent to bug injection research [13, 48, 49, 71]. Regarding Java, *LEAP* leverages bug injection to create a benchmark of Java concurrency bugs [23], and *Pan et al.* define 27 bug patterns for automatic program repair [45]. None of these works cover Java deserialization gadget chains as a bug type. *Ladisa et al.* injected malicious bytecode into benign JAR files to evaluate different malicious package detection strategies [31]. Instead of using immature vulnerabilities, the authors inserted three standalone payloads.

*Wu et al.* pointed out the high false alarm rate in supply chain dependents. They showed that upstream dependencies in the *Maven* ecosystem affect only *10.4%* of downstream projects through usage of a vulnerable upstream function [69]. However, this rate is not representative for our study since the insecure entry point ObjectInputStream.readObject() is assumed to be present in the downstream project while the choice of gadgets from the upstream project is unconstrained. The downstream project incorporates the gadgets in its classpath and is, thus, not required to explicitly use any vulnerable function. This means that if the gadget chain exists in a dependency, then it also exists within the user. There are ways to mitigate the impact through tools like *SBOM.EXE* [56], which restricts the set of permissible classes during runtime.

## 7 Conclusion
In this work, we investigated the usage of the Java *Serializable API* in a large dataset of 1 100 popular *Maven* dependencies over their evolution. It showed that in dependencies that use *Serializable*, the number of serializable classes has a weak positive correlation with time. This means, if anything, *Serializable* usage is increasing. From here, we defined four dependency datasets, which we propose as a baseline for future gadget chain detection research. Then, for the dependencies in these datasets, we injected three bug patterns to activate dormant gadget chains. We ran three gadget chain detection tools on both the modified and untouched dependencies and compared the results. For 139 dependencies, the tools detected additional gadget chains. This provides a soft proof of the patterns' viability to activate dormant gadget chains. Furthermore, we manually confirmed dormant gadget chains in 53 dependencies. It corresponds to a success rate of *9.94%* for our injection patterns.

Overall, our work showed that Java deserialization gadget chains are a much broader liability to software than detecting full gadget chains could possibly capture. It makes awareness of dormant gadget chains paramount to strengthen the software supply chain.

## 8 Data Availability
We anonymously share our experimentation data and toolchain with the following URL:

https://zenodo.org/records/15039856?preview=1&token=eyJhbGciOiJIUzUxMiIsImlh
dCI6MTc0MjIxODg4MiwiZXhwIjoxNzY3MTM5MTk5fQ.eyJpZCI6ImY4YzQ1ZDk
wLTIwOGQtNDEzMy1iODVlLWFiYTc5ODljODQxNiIsImRhdGEiOnt9LCJyYW5kb
20iOiJjZWIwMWZhMTA5N2U3NTU4ZWQ0MGVlMTA1OGE4OTIwNSJ9.iGAqYq
Gv1_MdFHnd8QcXtuS4mFO5zVvFLI78OyiKHqLD1phW5sYDkH_aTIrDAWximJK4
WmlzN0KFp_bVCM-xVQ

We do not disclose the gadget chains detected by the tools in the unmodified dependencies. It was not within this work's scope to verify these gadget chain detections, and, thus, by publishing the results, we would risk unintentionally disclosing real vulnerabilities.

## Acknowledgments

This work was partially supported by the Wallenberg AI, Autonomous Systems and Software Program (WASP) funded by the Knut and Alice Wallenberg Foundation.

# A  Maven Repository Download URL Construction

```
https://<repository>/<package>/<dependency>/<version>
    /<dependency>-<version>.<jar|aar>
# Example:
https://repo1.maven.org/maven2/org/slf4j/slf4j-api/1.1.0
    /slf4j-api-1.1.0.jar
```

Listing 12: General structure of *Maven* repository download URLs.

| |
|---|
| *https://repo1.maven.org/maven2* |
| *https://repo.clojars.org* |
| *https://repo.akka.io/maven* |
| *https://maven.google.com* |
| *https://maven.artifacts.atlassian.com* |
| *https://maven.wso2.org/nexus/content/repositories/releases* |
| *https://nexus.bedatadriven.com/content/groups/public* |
| *https://repository.mulesoft.org/nexus/content/repositories/public* |
| *https://repo.jenkins-ci.org/releases* |
| *https://nexus.senbox.net/nexus/content/repositories/releases* |
| *https://open.artefacts.tax.service.gov.uk/maven2* |

Table 8: Maven repository base URLs.

## B Sink Method Filtering for Gadget Chain Verification

| Sink Method | Count |
|---|---|
| java.lang.reflect.Method.invoke | 608 |
| java.lang.ClassLoader.defineClass | 32 |
| org.springframework.jndi.JndiTemplate.lookup | 3 |
| java.sql.PreparedStatement.execute | 1 |
| java.io.FileOutputStream.write | 168 |
| java.lang.ClassLoader.loadClass | 77 |
| java.net.URL.openConnection | 142 |
| java.sql.Statement.execute | 1 |
| javax.naming.InitialContext.lookup | 1 |
| java.lang.reflect.Constructor.newInstance | 423 |
| java.io.FileOutputStream.<init> | 91 |
| java.sql.PreparedStatement.executeQuery | 3 |
| java.io.File.delete | 91 |
| java.beans.Introspector.getBeanInfo | 3 |
| java.net.URL.openStream | 119 |
| java.sql.DriverManager.getConnection | 1 |
| java.sql.Connection.prepareStatement | 4 |
| java.nio.file.Files.newOutputStream | 28 |
| javax.naming.Context.lookup | 900 |
| java.lang.ProcessBuilder.<init> | 11 |
| java.lang.Runtime.exec | 18 |
| java.rmi.registry.Registry.lookup | 1 |
| java.nio.file.Files.newBufferedWriter | 7 |
| *Filtered* | |
| java.io.FileInputStream.<init> | 202 |
| java.lang.Class.forName | 239 |
| java.lang.Class.getMethod | 260 |
| java.lang.Class.getDeclaredMethod | 48 |
| java.lang.ClassLoader.getSystemResourceAsStream | 12 |
| java.lang.Class.getResourceAsStream | 5 |
| java.nio.file.Files.readAllLines | 7 |
| java.io.FileReader.<init> | 1 |
| org.xml.sax.XMLReader.parse | 4 |
| java.io.RandomAccessFile.read | 4 |
| java.io.RandomAccessFile.readFully | 2 |
| java.nio.file.Files.newInputStream | 81 |
| java.nio.file.Files.newBufferedReader | 5 |
| java.nio.file.Files.readAllBytes | 9 |
| java.util.zip.ZipInputStream.<init> | 1 |
| com.esotericsoftware.kryo.Kryo.readClassAndObject | 2 |

Table 9: Sink methods in dormant gadget chains. The sink methods in the lower section are filtered out for the manual verification. Note that we also disregard the dynamically identified sinks by *Crystallizer*. The definition of considering any method which operates on arbitrary objects produces a high amount of false positive sink methods [60].

## C Verified Dormant Gadget Chains
### C.1 Transitive Serializability

```
java.util.Hashtable.readObject()
↪ java.util.Hashtable.readHashtable()
↪ java.util.Hashtable.reconstitutionPut()
↪ com.github.jknack.handlebars.io.ReloadableTemplateSource.equals()
↪ com.github.jknack.handlebars.io.URLTemplateSourc.lastModified()
↪ com.github.jknack.handlebars.io.URLTemplateSource(URL)
↪ java.net.URL.openConnection()
```
Table 10: *com.github.jknack.handlebars-4.4.0*

### C.2 Final Attributes

```
org.apache.flink.api.common.state.StateDescriptor.readObject()
↪ org.apache.flink.api.java.typeutils.runtime.kryo.KryoSerializer.deserialize()
↪ org.apache.flink.api.java.typeutils.runtime.kryo.KryoSerializer.checkKryoInitialized()
↪ org.apache.flink.api.java.typeutils.runtime.kryo.KryoSerializer.getKryoInstance()
↪ java.lang.reflect.Method.invoke()
```
Table 11: *org.apache.flink.flink-core-1.19.2*

```
org.jfree.chart.plot.CombinedRangeCategoryPlot.readObject()
↪ org.jfree.chart.axis.PeriodAxis.configure()
↪ org.jfree.chart.axis.PeriodAxis.autoAdjustRange()
↪ org.jfree.chart.axis.PeriodAxis.createInstance()
↪ java.lang.reflect.Constructor.newInstance()
```
Table 12: *org.jfree.jfreechart-1.5.5*

```
org.rogach.scallop.SerializationProxy.readObject()
↪ java.lang.reflect.Constructor.newInstance()
```
Table 13: *org.rogach.scallop-5.2.0*

### C.3 Interface Method Reachability

```
java.lang.Runnable.run()
↪ org.springframework.beans.factory.support.DisposableBeanAdapter.run()
↪ org.springframework.beans.factory.support.DisposableBeanAdapter.destroy()
↪ org.springframework.beans.factory.support.DisposableBeanAdapter.invokeCustomDestroyMethod()
↪ java.lang.reflect.Method.invoke()
```
Table 14: *org.springframework.spring-beans-6.1.17*

```
java.beans.PropertyChangeListener.propertyChange()
↪ org.htmlparser.beans.HTMLTextBean.propertyChange()
↪ javax.swing.text.JTextComponent.setText()
↪ javax.swing.text.AbstractDocument.replace()
↪ javax.swing.text.DefaultFormatter$DefaultDocumentFilter.replace()
↪ javax.swing.text.NumberFormatter.replace()
↪ javax.swing.text.NumberFormatter.toggleSignIfNecessary()
↪ javax.swing.text.NumberFormatter.toggleSign()
↪ java.lang.reflect.Constructor.newInstance()
```
Table 15: *org.htmlparser.htmlparser-2.1*

```
javax.sql.DataSource.getConnection()
↪ org.apache.commons.dbcp2.datasources.InstanceKeyDataSource.getConnection()
↪ org.apache.commons.dbcp2.datasources.InstanceKeyDataSource.testCPDS()
↪ javax.naming.InitialContext.lookup()
↪ javax.naming.Context.lookup()
```
Table 16: *org.apache.commons.commons-dbcp2-2.13.0*

```
java.lang.AutoCloseable.close()
↪ org.mapdb.BTreeMap.close()
↪ org.mapdb.StoreTrivialTx.close()
↪ java.io.File.delete()
```
Table 17: *org.mapdb.mapdb-3.1.0*

```
java.lang.reflect.InvocationHandler.invoke()
↪ org.apache.openejb.threads.impl.ContextServiceImpl$CUHandler.invoke()
↪ org.apache.openejb.threads.impl.ContextServiceImpl$CUHandler.lambda$invoke$0()
↪ java.lang.reflect.Method.invoke()
```
Table 18: *org.apache.tomee.openejb-core-10.0.0*

```
java.lang.Runnable.run()
↪ com.googlecode.aviator.runtime.function.AbstractVariadicFunction.run()
↪ com.googlecode.aviator.runtime.function.AbstractVariadicFunction.call()
↪ com.googlecode.aviator.runtime.function.ClassMethodFunction.variadicCall()
↪ com.googlecode.aviator.utils.Reflector.invokeStaticMethod()
↪ com.googlecode.aviator.utils.Reflector.invokeMatchingMethod()
↪ java.lang.reflect.Method.invoke()
```
Table 19: *com.googlecode.aviator.aviator-5.4.2*

```
javax.sql.RowSet.rollback()
↪ oracle.jdbc.rowset.OracleCachedRowSet.rollback()
↪ oracle.jdbc.rowset.OracleCachedRowSet.getConnectionInternal()
↪ javax.naming.InitialContext.lookup()
↪ javax.naming.Context.lookup()
```
Table 20: *com.oracle.database.jdbc.ojdbcx-21.17.0.0* - this gadget chain is the same for ojdbc 11, 8 and 6

```
java.lang.reflect.InvocationHandler.invoke()
↪ org.hibernate.validator.internal.util.annotation.AnnotationProxy.invoke()
↪ java.lang.reflect.Method.invoke()
```
Table 21: *org.hibernate.validator.hibernate-validator-8.0.2*

```
java.util.concurrent.Callable.call()
↪ org.redisson.mapreduce.CoordinatorTask.call()
↪ java.lang.reflect.Constructor.newInstance()
```
Table 22: *org.redisson.redisson-3.44.0*

```
java.lang.Runnable.run()
↪ weka.gui.experiment.RunPanel$ExperimentRunner.run()
↪ weka.experiment.Experiment.initialize()
↪ weka.experiment.RandomSplitResultProducer.preProcess()
↪ weka.experiment.CSVResultListener.preProcess()
↪ java.io.FileOutputStream.<init>()
```
Table 23: *nz.ac.waikato.cms.weka.weka-dev-3.9.6*

```
java.lang.reflect.InvocationHandler.invoke()
↪ org.apache.bval.jsr.util.AnnotationProxy.invoke()
↪ java.lang.reflect.Method.invoke()
```
Table 24: *org.apache.bval.bval-jsr-3.0.1*

```
java.awt.event.ActionListener.actionPerformed()
↪ org.jdesktop.swingx.action.ServerAction.actionPerformed()
↪ java.net.URL.openConnection()
```
Table 25: *org.swinglabs.swingx-1.6.1*

```
java.lang.reflect.InocationHandler.invoke()
↪ org.apache.ibatis.logging.jdbc.ResultSetLogger.invoke()
↪ java.lang.reflect.Method.invoke()
```
Table 26: *org.mybatis.mybatis-3.5.19*

```
javax.sql.XADataSource.getXAConnection()
↪ com.alibaba.druid.pool.xa.getXAConnection()
↪ com.alibaba.druid.pool.DruidDataSource.getConnection()
↪ com.alibaba.druid.pool.DruidDataSource.getConnection()
↪ com.alibaba.druid.pool.DruidDataSource.getConnectionDirect()
↪ com.alibaba.druid.pool.DruidDataSource.getConnectionInternal()
↪ com.alibaba.druid.pool.DruidDataSource.takeLast()
↪ com.alibaba.druid.pool.DruidDataSource.pollLast()
↪ com.alibaba.druid.pool.DruidDataSource$CreateConnectionThread.run()
↪ com.alibaba.druid.pool.DruidAbstractDataSource.createPhysicalConnection()
↪ com.alibaba.druid.pool.DruidAbstractDataSource.createPhysicalConnection()
↪ java.sql.Driver.connect()
```
Table 27: *com.alibaba.druid-1.2.24*

```
java.lang.reflect.InvocationHandler.invoke()
↪ org.springframework.core.SerializableTypeWrapper$TypeProxyInvocationHandler.invoke()
↪ org.springframework.util.ReflectionUtils.invokeMethod()
↪ java.lang.reflect.Method.invoke()
```
Table 28: *org.springframework.spring-core-6.1.17*

```
javax.naming.Context.lookup()
↪ org.apache.activemq.jndi.ReadOnlyContext.lookup()
↪ javax.naming.Context.lookup()
```
Table 29: *org.apache.activemq.activemq-client-6.1.5*. The interesting part is that `ReadOnlyContext` is serializable

```
javax.naming.Context.lookup()
↪ org.apache.camel.util.jndi.JndiContext.lookup()
↪ javax.naming.Context.lookup()
```
Table 30: *org.apache.camel.camel-core-2.25.4*

```
javax.sql.DataSource.getConnection()
↪ net.sourceforge.jtds.jdbcx.JtdsDataSource.getConnection()
↪ net.sourceforge.jtds.jdbcx.JtdsDataSource.getConnection(String, String)
↪ java.io.FileOutputStream.<init>()
```
Table 31: *net.sourceforge.jtds.jtds-1.2.8*

```
java.lang.reflect.InvocationHandler.invoke()
↪ io.lettuce.core.dynamic.support.TypeWrapper$TypeProxyInvocationHandler.invoke()
↪ java.lang.reflect.Method.invoke()
```
Table 32: *io.lettuce.lettuce-core-6.5.4*

```
javax.sql.DataSource.getConnection()
↪ org.apache.derby.client.BasicClientDataSource.getConnection()
↪ org.apache.derby.client.BasicClientDataSource.computeDncLogWriterForNewConnection()
↪ org.apache.derby.client.BasicClientDataSource.computeDncLogWriterForNewConnection()
↪ org.apache.derby.client.BasicClientDataSource.computeDncLogWriter()
↪ org.apache.derby.client.BasicClientDataSource.computePrintWriter()
↪ org.apache.derby.client.BasicClientDataSource.getPrintWriter()
↪ java.io.FileOutputStream.<init>()
```
Table 33: *org.apache.derby.derbyclient-10.17.1.0*

```
javax.xml.transform.Transformer newTransformer()
↪ com.sun.org.apache.xalan.internal.xsltc.trax.TemplatesImpl.newTransformer()
↪ com.sun.org.apache.xalan.internal.xsltc.trax.TemplatesImpl.getTransletInstance()
↪ java.lang.reflect.Constructor.newInstance()
```
Table 34: *xalan.xalan-2.7.3*

## C.4 Mixed

```
java.lang.Runnable.run()
↪ org.apache.hadoop.security.UserGroupInformation$AutoRenewalForUserCredsRunnable.run()
↪ org.apache.hadoop.security.UserGroupInformation$TicketCacheRenewalRunnable.relogin()
↪ org.apache.hadoop.util.Shell.execCommand(java.lang.String[])
↪ org.apache.hadoop.util.Shell.execCommand(java.util.Map, java.lang.String[], long)
↪ org.apache.hadoop.util.Shell$ShellCommandExecutor.execute()
↪ org.apache.hadoop.util.Shell.run()
↪ org.apache.hadoop.util.Shell.runCommand()
↪ java.lang.ProcessBuilder.<init>()
```
Table 35: *org.apache.hadoop.hadoop-common-3.4.1*

```
java.util.Hashtable.readObject()
↪ java.util.Hashtable.readHashtable()
↪ java.util.Hashtable.reconstitutionPut()
↪ java.net.URL.equals()
↪ org.eclipse.osgi.internal.url.MultiplexingURLStreamHandler.equals()
↪ org.eclipse.osgi.internal.url.MultiplexingURLStreamHandler.findAuthorizedURLStreamHandler()
↪ org.eclipse.osgi.internal.url.URLStreamHandlerFactoryImpl.findAuthorizedURLStreamHandler()
↪ java.lang.reflect.Method.invoke()
```
Table 36: *org.eclipse.platform.org.eclipse.osgi-3.22.0*

```
java.lang.AutoCloseable.close()
↪ java.util.Hashtable.readHashtable()
↪ org.apache.poi.openxml4j.opc.OPCPackage.close()
↪ org.apache.poi.openxml4j.opc.ZipPackage.closeImpl()
↪ org.apache.poi.openxml4j.opc.OPCPackage.save()()
↪ java.nio.file.Files.newOutputStream()
```
Table 37: *org.apache.poi.poi-ooxml-5.4.0*

```
java.awt.event.ActionListener.actionPerformed()
↪ org.h2.tools.GUIConsole.actionPerformed()
↪ org.h2.tools.GUIConsole.shutdown()
↪ org.h2.util.Utils.callMethod()
↪ org.h2.util.Utils.callMethod()
↪ java.lang.reflect.Method.invoke()
```
Table 38: *com.h2database.h2-2.3.232*

```
java.lang.AutoCloseable.close()
↪ org.apache.hadoop.fs.store.DataBlocks$DataBlock.close()
↪ org.apache.hadoop.fs.store.DataBlocks$DiskBlock.innerClose()
↪ org.apache.hadoop.fs.store.DataBlocks$DiskBlock.closeBlock()
↪ java.io.File.delete()
```
Table 39: *org.apache.hadoop.hadoop-common-3.4.1*

```
java.lang.AutoCloseable.close()
↪ org.apache.cassandra.io.util.RewindableDataInputStreamPlus.close()
↪ org.apache.cassandra.io.util.RewindableDataInputStreamPlus.close(boolean)
↪ java.io.File.delete()
```
Table 40: *org.apache.cassandra.cassandra-all-3.0.32*

```
java.lang.Runnable.run()
↪ org.apache.logging.log4j.core.appender.rolling.action.AbstractAction.run()
↪ org.apache.logging.log4j.core.appender.rolling.action.ZipCompressAction.execute()
↪ org.apache.logging.log4j.core.appender.rolling.action.ZipCompressAction.execute()
↪ java.io.File.delete()
```
Table 41: *org.apache.log4j.log4j-core-2.24.3*

```
java.util.Iterator.hasNext()
↪ org.apache.openjpa.jdbc.meta.strats.LRSProxyMap$ResultIterator.hasNext()
↪ org.apache.openjpa.jdbc.sql.MergedResult.next()
↪ org.apache.openjpa.jdbc.sql.LogicalUnion$ResultComparator.getOrderingValue(Result, int)
↪ org.apache.openjpa.jdbc.sql.LogicalUnion$ResultComparator.getOrderingValue(Result, Object)
↪ org.apache.openjpa.jdbc.sql.PostgresDictionary.getObject()
↪ java.lang.reflect.Method.invoke()
```
Table 42: *org.apache.openjpa.openjpa-4.0.1*

```
java.lang.Runnable.run()
↪ org.jgroups.protocols.MERGE3$InfoSender.run()
↪ org.jgroups.protocols.PDC.down()
↪ org.jgroups.protocols.PDC.writeNodeToDisk()
↪ org.jgroups.protocols.PDC.writeToTempFile()
↪ java.io.FileOutputStream.<init>()
```
Table 43: *org.jgroups.jgroups-5.3.15*

```
java.beans.PropertyChangeListener.propertyChange()
↪ org.apache.catalina.core.NamingContextListener.propertyChange()
↪ org.apache.catalina.core.NamingContextListener.processGlobalResourcesChange()
↪ org.apache.catalina.core.NamingContextListener.addEnvironment()
↪ org.apache.catalina.core.NamingContextListener.constructEnvEntry()
↪ java.lang.reflect.Constructor.newInstance()
```
Table 44: *org.apache.tomcat.embed.tomcat-embed-core-10.1.36*
and *org.apache.tomcat.tomcat-catalina-10.1.36*

```
java.util.Iterator.hasNext()
↪ org.python.core.WrappedIterIterator.hasNext()
↪ org.python.modules.itertools.imap$1.__iternext__()
↪ org.python.core.PyBuiltinMethodNarrow.__call__()
↪ org.python.core.PyJavaType$14.__call__()
↪ java.lang.reflect.Method.invoke()
```
Table 45: *org.python.jython-standalone-2.7.4*

```
java.lang.Runnable.run()
↪ net.bytebuddy.ClassFileVersion$VersionLocator$Resolver.run()
↪ java.lang.reflect.Method.invoke()
```
Table 46: *net.bytebuddy.byte-buddy-1.17.1*

```
java.util.Map.put()
↪ org.apache.commons.beanutils.BaseDynaBeanMapDecorator.put()
↪ org.apache.commons.beanutils.LazyDynaBean.get()
↪ org.apache.commons.beanutils.LazyDynaBean.createProperty()
↪ org.apache.commons.beanutils.LazyDynaBean.createOtherProperty()
↪ java.lang.reflect.Constructor.newInstance()
```
Table 47: *commons-beanutils-1.10.1*

```
java.io.Flushable.flush()
↪ org.springframework.integration.metadata.PropertiesPersistingMetadataStore.flush()
↪ org.springframework.integration.metadata.PropertiesPersistingMetadataStore.saveMetaData()
↪ java.io.FileOutputStream.<init>()
```
Table 48: *spring-integration-core-6.3.8*

```
java.util.Iterator.next()
↪ com.mysql.cj.xdevapi.AbstractDataResult.next()
↪ com.mysql.cj.protocol.ProtocolEntityFactory.createFromProtocolEntity()
↪ com.mysql.cj.exceptions.ExceptionFactory.createException()
↪ java.lang.reflect.Constructor.newInstance()
```
Table 49: *com.mysql.mysql-connector-j-9.2.0*

```
java.lang.AutoCloseable.close()
↪ org.apache.sshd.agent.unix.AgentServerProxy.close()
↪ org.apache.sshd.agent.unix.AgentServerProxy.removeSockerFile()
↪ org.apache.sshd.agent.unix.AgentServerProxy.deleteFile()
↪ java.io.File.delete()
```
Table 50: *org.apache.sshd.sshd-core-2.14.0*

```
java.lang.Runnable.run()
↪ org.apache.calcite.adapter.druid.DruidConnectionImpl$1$1.run()
↪ org.apache.calcite.adapter.druid.DruidConnectionImpl.request()
↪ org.apache.calcite.runtime.HttpUtils.post()
↪ org.apache.calcite.runtime.HttpUtils.executeMethod()
↪ org.apache.calcite.runtime.HttpUtils.getURLConnection()
↪ java.net.URL.openConnection()
```
Table 51: *org.apache.hive.hive-exec-4.0.1*

```
java.lang.reflect.InvocationHandler.invoke()
↪ org.codehaus.groovy.runtime.ConversionHandler.invoke()
↪ java.lang.reflect.Method.invoke()
```
Table 52: *org.apache.groovy.groovy-4.0.25*

```
java.lang.Iterable.iterator()
↪ com.google.javascript.jscomp.jarjar.org.apache.tools.ant.types.resources.Files.iterator()
↪ com.google.javascript.jscomp.jarjar.org.apache.tools.ant.types.ArchiveScanner.scan()
↪ com.google.javascript.jscomp.jarjar.org.apache.tools.ant.types.resources.URLResource.isExists()
↪ com.google.javascript.jscomp.jarjar.org.apache.tools.ant.types.resources.URLResource.isExists()
↪ ↪ com.google.javascript.jscomp.jarjar.org.apache.tools.ant.types.resources.URLResource.connect()
↪ java.net.URL.openConnection()
```
Table 53: *com.google.javascript.closure-compiler-20240317*

```
java.lang.Runnable.run()()
↪ io.undertow.server.handlers.resource.URLResource$1ServerTask.run()
↪ java.net.URL.openStream()
```
Table 54: *io.undertow.undertow-core-2.2.37*

```
java.lang.Runnable.run()()
↪ io.undertow.server.handlers.resource.URLResource$1ServerTask.run()
↪ java.net.URL.openStream()
```
Table 55: *io.undertow.undertow-core-2.2.37*

```
java.sql.Connection.isValid()
↪ com.clickhouse.jdbc.internal.ClickHouseConnectionImpl.isValid()
↪ com.clickhouse.client.AbstractClient.ping()
↪ com.clickhouse.client.http.ClickHouseHttpClient.checkHealth()
↪ com.clickhouse.client.http.HttpUrlConnectionImpl.ping()
↪ com.clickhouse.client.http.HttpUrlConnectionImpl.newConnection()
↪ java.net.URLConnection.openConnection()
```
Table 56: *com.clickhouse.clickhouse-jdbc-0.8.1*

```
java.lang.AutoCloseable.close()
↪ liquibase.database.core.DerbyDatabase.close()
↪ liquibase.database.core.DerbyDatabase.shutdownDerby()
↪ java.lang.reflect.Constructor.newInstance()
```
Table 57: *org.liquibase.liquibase-core-4.31.1*

```
java.util.Iterator.next()
↪ edu.stanford.nlp.objectbank.ReaderIteratorFactory$ReaderIterator.next()
↪ edu.stanford.nlp.objectbank.ReaderIteratorFactory$ReaderIterator.setNextObject()
↪ java.net.URL.openStream()
```
Table 58: *edu.stanford.nlp.stanford-corenlp-4.5.8*

```
java.util.Iterator.hasNext()
↪ org.eclipse.persistence.queries.ScrollableCursor.hasNext()
↪ org.eclipse.persistence.queries.ScrollableCursor.loadNext()
↪ org.eclipse.persistence.queries.ScrollableCursor.retrieveNextObject()
↪ org.eclipse.persistence.internal.queries.JoinedAttributeManager.processDataResults()
↪ org.eclipse.persistence.internal.descriptors.ObjectBuilder.extractPrimaryKeyFromRow()
↪ org.eclipse.persistence.internal.descriptors.ObjectBuilder.extractPrimaryKeyFromObject()
↪ org.eclipse.persistence.internal.descriptors.ObjectBuilder.extractPrimaryKeyFromObject()
↪ org.eclipse.persistence.mappings.structures.NestedTableMapping.writeFromObjectIntoRow()
↪ org.eclipse.persistence.internal.databaseaccess.DatasourceAccessor.incrementCallCount()
↪ org.eclipse.persistence.internal.databaseaccess.DatasourceAccessor.reconnect()
↪ org.eclipse.persistence.internal.databaseaccess.DatasourceAccessor.connectInternal()
↪ org.eclipse.persistence.sessions.DatasourceLogin.connectToDataSource()
↪ org.eclipse.persistence.sessions.JNDIConnector.connect()
↪ javax.naming.Context.lookup()
```
Table 59: *org.eclipse.persistence.eclipselink-4.0.5*

```
java.util.Iterator.hasNext()()
↪ org.hibernate.query.internal.ScrollableResultsIterator.hasNext()
↪ org.hibernate.internal.FetchingScrollableResultsImpl.next()
↪ org.hibernate.internal.FetchingScrollableResultsImpl.prepareCurrentRow()
↪ org.hibernate.sql.results.internal.StandardRowReader.readRow()
↪ org.hibernate.sql.results.internal.RowTransformerTupleTransformerAdapter.transformRow()
↪ org.hibernate.jpa.spi.NativeQueryConstructorTransformer.transformTuple()
↪ java.lang.reflect.Constructor.newInstance()
```
Table 60: *org.hibernate.orm.hibernate-core-6.6.8*